\documentclass{article}
\usepackage{spconf,amsmath,graphicx,hyperref}

\usepackage{tipa} 
\usepackage{svg}
\usepackage{booktabs}   

\title{voice2mode: Phonation Mode Classification in Singing using Self-Supervised Speech Models}
\name{
Aju Ani Justus$^{1}$,
Ruchit Agrawal$^{2}$,
Sudarsana Reddy Kadiri$^{3}$,
Shrikanth Narayanan$^{3}$
}

\address{
$^{1}$University of Birmingham, School of Computer Science, Birmingham, UK \\
$^{2}$Carnegie Mellon University, Information Systems, Doha, Qatar \\
$^{3}$University of Southern California, Department of Electrical and Computer Engineering, LA, USA
}
\begin{document}
\ninept
\maketitle
\begin{abstract}
We present \textbf{voice2mode}, a method for classification of four singing phonation modes (breathy, neutral (modal), flow, and pressed) using embeddings extracted from large self-supervised speech models. Prior work on singing phonation has relied on handcrafted signal features or task-specific neural nets; this work evaluates the transferability of speech foundation models to singing phonation classification. \textbf{voice2mode} extracts layer-wise representations from HuBERT and two wav2vec2 variants, applies global temporal pooling, and classifies the pooled embeddings with lightweight classifiers (SVM, XGBoost). Experiments on a publicly available soprano dataset (763 sustained vowel recordings, four labels) show that foundation-model features substantially outperform conventional spectral baselines (spectrogram, mel-spectrogram, MFCC). HuBERT embeddings obtained from early layers yield the best result ($\approx$95.7\% accuracy with SVM), an absolute improvement of $\approx$12–15\% over the best traditional baseline. We also show layer-wise behaviour: lower layers, which retain acoustic/phonetic detail, are more effective than top layers specialized for Automatic Speech Recognition (ASR).
\end{abstract}

\begin{keywords}
phonation mode classification; singing; self-supervised models; speech and audio processing; transfer learning.
\end{keywords}

\section{Introduction}\label{sec:introduction}
The automatic classification of singing \textbf{phonation modes} (voice qualities), such as \textit{breathy, neutral/modal, flow, and pressed}, plays a key role in vocal training and music analysis \cite{sopranodataproutskova2013,brandner2023classification}. Each phonation mode is associated with specific vocal-fold tension and glottal configurations. For example, \textit{breathy} singing has minimal glottal closure (a persistent glottal gap producing turbulent noise) \cite{sopranodataproutskova2013}, \textit{modal} (neutral) voice involves full vibration and complete closure, \textit{pressed} voice is marked by high subglottal pressure and tense adduction \cite{sund,vq_correlates}, and \textit{flow} (resonant) voice lies in between these extremes. 

Previous phonation-mode classification studies have largely relied on hand-crafted signal features (e.g. glottal inverse-filtering, cepstral measures, modulation spectra) and conventional classifiers \cite{brandner2023classification, kadiri2020analysis}. 
In recent years, \textbf{self-supervised foundation models} pre-trained on large speech or audio corpora, such as \textit{wav2vec2.0} and \textit{HuBERT}, have revolutionized speech processing. These models learn high-level representations of raw audio by predicting masked segments or hidden units, and their embeddings have yielded state-of-the-art results in tasks like speech recognition, emotion detection, speaker identification, and voice-quality classification \cite{hsu2021hubert, aly2021, KADIRI2024101550}. While self-supervised learning speech models have proven effective for speech voice-quality classification, their applicability to singing phonation remains under-explored. We therefore investigate whether speech-based foundation models can extract meaningful features from singing, conditioned on the simplified phonation labels provided by the reference dataset.

We present \textbf{voice2mode}, the first phonation-mode classifier for singing that uses embeddings from pre-trained speech models. We consider three well-known self-supervised models as feature extractors: \textbf{HuBERT}, \textbf{wav2vec2.0-Base}, and \textbf{wav2vec2.0-Large}. Each model processes the raw singing waveform to produce layer-wise representations: we take the \textit{temporal average} of the output of each transformer layer (as well as the input “layer 0”) to form fixed-size feature vectors. These embeddings are then fed into simple classifiers (Support Vector Machine or XGBoost) to predict one of the four modes. In contrast to prior methods (which use engineered spectral or glottal features), voice2mode leverages the deep acoustic abstractions learned by foundation models. Our experiments use the publicly available soprano singing dataset \cite{sopranodataproutskova2013}. It consists of sustained vowels labelled as breathy, neutral, flow, or pressed. We compare voice2mode against conventional features (spectrogram, mel-spectrogram, MFCC) under a 5-fold cross-validation. The results show that embeddings from all three foundation models dramatically outperform the baselines. In particular, HuBERT-based features achieve around 95\% accuracy (with SVM), versus $\approx$80\% for the best conventional feature \cite{brandner2023classification}. HuBERT embeddings lead, followed by wav2vec2.0-Large and wav2vec2.0-Base. Notably, we find that \textbf{lower layers} of these models (which retain more phoneme-like acoustic information) yield the highest classification accuracy, while deeper layers (more specialized for ASR) are less effective. Overall, voice2mode improves absolute accuracy by roughly 12–15\% over baseline features; a substantial gain supported by prior findings that data-driven features can boost phonation-mode classification \cite{brandner2023classification, KADIRI2024101550}. Our key contributions are as follows:
\begin{itemize}
    \item We introduce voice2mode, the first singing voice phonation-mode classifier built on \textbf{speech foundation models} (HuBERT, wav2vec2.0-Base, wav2vec2.0-Large). This shows that speech-pre-trained self-supervised learning (SSL) models can generalize to singing tasks.
    \item We analyse layer-wise embeddings: for each model, we extract features from every layer and evaluate which layers best separate the four singing phonation modes.
    \item We benchmark these self-supervised learning (SSL) features against standard baselines (spectrogram, mel, MFCC) using both SVM and XGBoost classifiers. Our results demonstrate clear superiority of the SSL-based features.
    \item We provide a reproducible codebase to foster further research in singing voice analysis.\footnote{\url{https://github.com/ajuanijustus/voice2mode}}
\end{itemize}

\section{Related work}
Previous methods for automatic phonation-mode classification in singing have largely relied on classical signal processing techniques \cite{brandner2023classification, kadiri2020analysis, rouas2016automatic}. For example, \cite{brandner2023classification} proposed modulation spectral features for phonation-mode classification, while \cite{stoller2016analysis} investigated multiple acoustic features using a feed-forward neural network, showing that cepstral peak prominence (CPP), temporal flatness, and energy best separated phonation modes. Although these results demonstrate the potential of learned representations, they rely on shallow neural architectures and handcrafted features.

Beyond singing-specific studies, phonation modes such as breathy, whispery, creaky, and modal voices have been extensively studied in speech processing. A recent systematic review \cite{patman2025breathy} surveys acoustic approaches for analysing non-pathological voice qualities, highlighting the diversity of representations proposed for breathy and whispery phonation. A recent study \cite{kadiri2020analysis} showed that features originally developed for speech often perform suboptimally for singing, particularly at high pitches where glottal inverse filtering becomes unreliable. The authors therefore proposed advanced signal-processing features that avoid explicit source–filter separation.

Speech-oriented tools for explicit phonation manipulation have also been proposed, such as VoiceQualityVC \cite{lameris2025voicequalityvc}, an open-source voice conversion framework designed to study perceptual effects of voice quality in speech; however, such systems are not tailored to the acoustic characteristics of singing. With the rise of data-driven methods, several deep-learning approaches for singing phonation-mode detection have emerged \cite{sun2020residual, wang2023phonation, wang2023disentangled}. These include residual attention networks \cite{sun2020residual} and convolutional–recurrent encoder–decoder architectures \cite{wang2023phonation}, as well as disentangled domain adaptation techniques to handle label scarcity \cite{wang2023disentangled}. More recently, alternative representations such as wavelet scattering features have been explored for singing phonation classification, demonstrating improved robustness over conventional spectral features \cite{mittapalle2024wavelet}.

The latest advances in self-supervised learning have led to powerful speech foundation models such as HuBERT \cite{hsu2021hubert} and wav2vec~2.0 \cite{baevski2020wav2vec2}, which learn rich acoustic representations via masked prediction objectives and achieve strong performance in speech recognition. Recent surveys and benchmarks have further consolidated their role in speech processing \cite{cui2025speechlm, yang2024superb}, with growing interest in integrating speech foundation models with large language models \cite{verdini2025connect}.

Recently, \cite{KADIRI2024101550} demonstrated that HuBERT and wav2vec~2.0 embeddings significantly outperform conventional spectro-temporal features for speech voice-quality classification. However, the application of such foundation models to singing voice, particularly for phonation-mode classification, remains largely unexplored. Singing differs acoustically from speech due to sustained phonation, vibrato, and wide pitch ranges, while sharing the same underlying vocal mechanisms. The present study therefore investigates whether speech-based foundation models can generalize to singing phonation analysis, conditioned on the simplified phonation labels provided by the reference dataset.


\section{\textbf{voice2mode}: Phonation Mode Classification Framework}
The framework for phonation mode classification developed in this study has two main components, the Feature Extractor (FE) and the Phonation Mode Classifier (PMC), as illustrated in Figure 1. The FE component employs three well-known self-supervised pre-trained models (wav2vec2-BASE \cite{baevski2020wav2vec2}, wav2vec2-LARGE \cite{baevski2020wav2vec2}, and HuBERT \cite{hsu2021hubert} to extract feature vectors from singing voices. For the classifier, we experiment with Support Vector Machine (SVM) and XGBoost (XGB) classifiers. 


\begin{figure*}[t]
    \centering
    \includesvg[width=\textwidth]{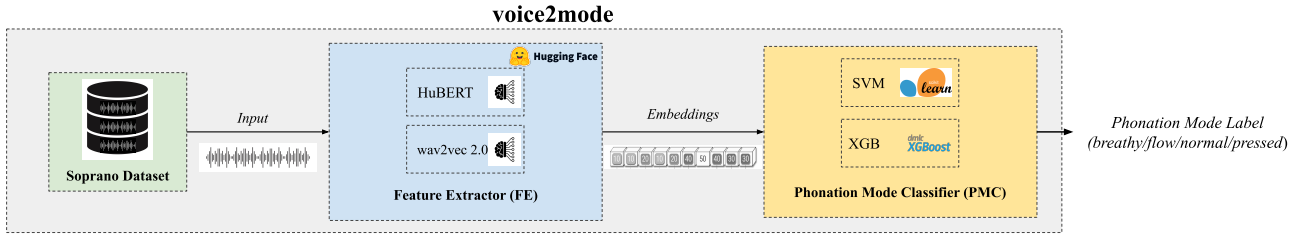}
    \caption{Schematic block diagram illustrating the proposed \textbf{voice2mode} phonation mode classification system for singing voice. The system utilizes foundation models (HuBERT and wav2vec2.0) as feature extractor, and SVM and XGBoost as classifiers. The raw signals are used directly as input to the Feature Extractor component that is based on pre-trained self-supervised models (wav2vec2 and HuBERT), and the two classifiers (SVM and XBG).}
    \label{fig:framework}
\end{figure*}

\subsection{Feature Extractor (FE)}
In this study, we use three pre-trained models as feature extractors:
\begin{itemize}
    \item wav2vec2.0-BASE: 12-layer Transformer encoder (768-dimensional each) with a convolutional feature encoder.
    \item wav2vec2.0-LARGE: 24-layer Transformer (1024-dimensional each).
    \item HuBERT: 24-layer Transformer (1024-dimensional), trained with masked prediction of clustered speech units.
\end{itemize}
All models were pre-trained on large English speech datasets (LibriSpeech, etc.) and fine-tuned for ASR. To extract features, each normalized singing waveform is fed into the model, and we record the output of every encoder layer (plus the raw CNN output, “layer 0”). We then collapse each layer’s sequence of hidden vectors into a single fixed vector by averaging over time (i.e. global mean pooling). This yields one 768-D feature for layer 0 and 768-D for each of 12 layers in wav2vec-base (13 vectors total), 1024-D for each of 24 layers in wav2vec-large (25 vectors total), and similarly 25 vectors of 1024-D from HuBERT. In other words, we produce a hierarchy of representations for each audio example. Prior work has shown that lower layers capture general acoustic/phone information, while higher layers are more linguistic \cite{brandner2023classification, KADIRI2024101550} . We evaluate each layer’s features separately, as well as collectively, to see which best discriminate phonation modes.

\subsection{Phonation Mode Classifier (PMC)}
With the extracted features, we train two types of classifiers: a linear-kernel Support Vector Machine (SVM) and an eXtreme Gradient Boosting (XGBoost) classifier. We tune hyperparameters (C, gamma for SVM; learning rate, depth, etc. for XGB) via grid search with 5-fold cross-validation. Due to limited data size (see Section~\ref{database} for more details), we keep the classifiers relatively simple. All evaluations use stratified 5-fold CV (each fold containing all four classes) to compute average accuracy and confusion matrices.

\section{Experimental Setup}
This section provides a description of the experimental setup, including the singing voice database, baseline features, the training and testing strategy along with the evaluation metrics.

\subsection{Singing voice database}
\label{database}
We use a freely and publicly available soprano database, consisting of singing voices showcasing various phonation modes. This database contains sustained vowels performed by a professional female Russian vocalist \cite{sopranodataproutskova2013}. The phonation modes present in the database are: breathy, modal, flow, and pressed voice \footnote{The dataset is available under a Creative Commons license at http://www.proutskova.de/phonation-modes/}. After pre-screening the data, a total of 763 recordings across nine distinct vowels are available in the following vowel categories: {{\textipa{A}}, {\textipa{AE}}, {\textipa{I}}, {\textipa{O}}, {\textipa{U}}, {\textipa{UE}}, {\textipa{Y}}, {\textipa{OE}}, and \textipa{E}}, with pitches spanning from A3 to G5. Recordings were conducted at a sampling frequency of 44.1~kHz. Further elucidation on the databases is provided by Proutskova et al. \cite{sopranodataproutskova2013}. 
To ensure uniformity for classification purposes, the database was down-sampled to 16~kHz and all the signals are normalized (range: -1 to +1). 

\subsection{Baseline features for comparative analysis}
Baseline Features: For comparison we also compute standard spectro-temporal features per recording: (a) \textit{Spectrogram}: 25ms Hamming windows, 5ms hop, 1024-FFT, log-magnitude spectrum averaged over time (513-D); (b) \textit{Mel-spectrogram}: 80 mel-bands, log-scale, averaged over time (80-D); (c) \textit{MFCCs}: 13 cepstral coeffs + deltas + double-deltas, averaged (39-D). These are common baselines in voice quality work \cite{brandner2023classification}.





\subsection{Training, testing and evaluation}
The classification experiments were conducted by employing a k-fold stratified cross-validation approach. Here, the data for each phonation mode was partitioned into 5 folds, with the SVM and XGB classifiers being trained iteratively on 4 folds while the fifth fold served as the test data. For each iteration, the classification accuracy was computed for the testing fold. This process was repeated 5 times, with 1 fold designated for testing in each iteration, totaling 5 iterations of training and testing. The final classification accuracy result was obtained by averaging over all iterations. Additionally, confusion matrices were computed to evaluate the class-wise performance.

\section{Results and Discussion}

This section presents the results and a comparative analysis of foundation model features juxtaposed with baseline features. 
Table \ref{table:conv_accuracy} displays the results (mean and standard deviation in accuracy) obtained from the conventional baseline features using the SVM and XGB classifiers. It can be observed that spectrogram feature performed the best among the baseline features. Between the classifiers, SVM performed better than XGB.

\begin{table}[ht]
\caption{Accuracy percentages (mean and standard deviation) for the three conventional features using the SVM and XGB classifiers. Best result highlighted in bold, second best result underlined.}
\vspace{-0.3cm}
\begin{tabular}[t]{p{3cm} p{1.5cm} p{1.5cm}}
\toprule
\textbf{Feature}&\textbf{SVM}&\textbf{XGB}\\
\midrule
Spectrogram&\textbf{79.9 ± 2.6}&79.6 ± 3.1\\
Mel-spectrogram&79.0 ± 2.7&\underline{79.8 ± 2.6}\\
MFCC&73.2 ± 4.2&74.1 ± 3.3\\
\midrule
\end{tabular}
\label{table:conv_accuracy}
\end{table}

\begin{table}[ht]
\caption{Accuracy percentages (mean and standard deviation) for the best-performing layer of foundation model features using the SVM and XGB classifiers. Best result highlighted in bold, second best result underlined.}
    \vspace{-0.3cm}
\begin{tabular}[t]{p{3cm} p{1.5cm} p{1.5cm}}
\toprule
\textbf{Feature}&\textbf{SVM}&\textbf{XGB}\\
\midrule
Spectrogram \newline (best baseline) & 79.9 ± 2.6 \newline & 79.6 ± 3.1\\  \hline 
wav2vec2-BASE&90.7 ± 5.1&83.7 ± 7.0\\
wav2vec2-LARGE&90.2 ± 5.3&82.6 ± 5.1\\
HuBERT&\textbf{95.7 ± 3.0}&\underline{92.0 ± 4.0}\\  
\midrule
\end{tabular}
\label{table:model_accuracy}
\end{table}

\begin{figure}[ht]
    \centering
\includesvg[width=1.1\columnwidth]{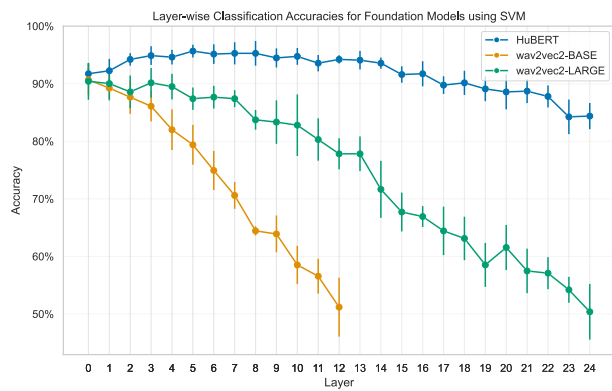}
 \vspace{-0.6cm}
\caption{Layer-wise classification accuracies for the features derived from the three pre-trained models: wav2vec2-BASE, wav2vec2-LARGE and HuBERT using SVM classifier.}
\label{fig:accuracy_comp1}
 \vspace{-0.5cm}

\end{figure}

\begin{figure}[ht]
    \centering
\includesvg[width=1.1\columnwidth]{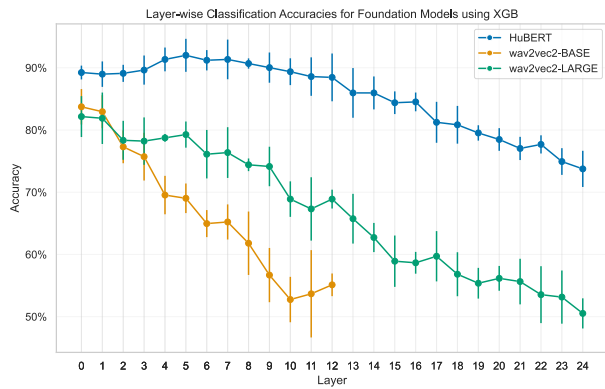}
\vspace{-0.6cm}
\caption{Layer-wise classification accuracies for the features derived from the three pre-trained models: wav2vec2-BASE, wav2vec2-LARGE and HuBERT using XGB classifier.}
    \label{fig:accuracy_comp2}
     \vspace{-0.5cm}
\end{figure}

\begin{figure}[ht]
    \centering
    \includegraphics[width=\columnwidth]{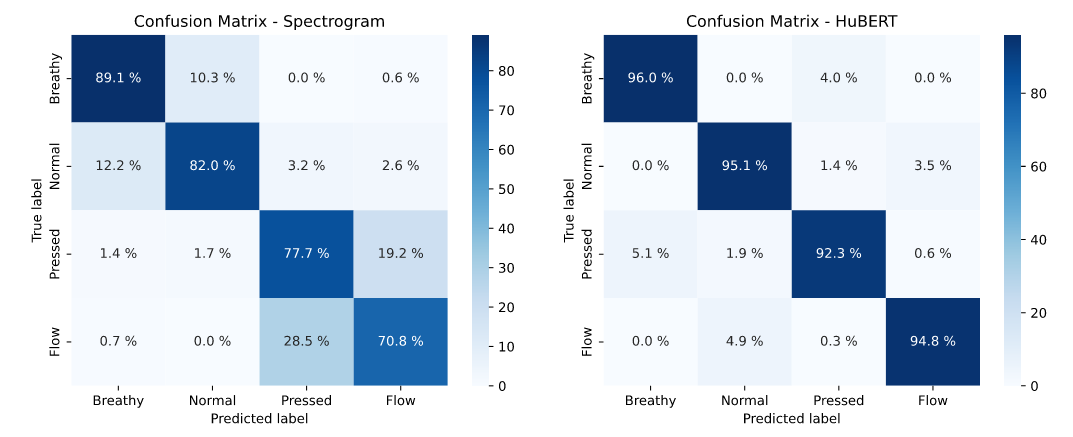}
        \vspace{-0.6cm}
    \caption{Confusion matrices for the best-performing baseline spectrogram feature (left) and for the best performing layer of HuBERT feature (right) using the SVM classifier.}
    \label{fig:cm}
            \vspace{-0.2cm}
\end{figure}

Figures \ref{fig:accuracy_comp1} and \ref{fig:accuracy_comp2} illustrate the layer-wise results (with mean and standard deviation in accuracy) for the features derived from the three pre-trained foundation models (HuBERT, wav2vec2-LARGE, and wav2vec2-BASE) using SVM and XGB classifier, respectively. Figures suggest that the features from all pre-trained foundation models, particularly those from the initial layers, outperformed the conventional features. Across all pre-trained models, it's evident that features from the initial layers exhibit superior performance compared to those from later layers, indicating the effectiveness of these initial layers in capturing generic features for phonation mode classification of singing voices. This behaviour is consistent with prior findings \cite{javanmardi2024exploring,javanmardi2024pre,vaidya2022ssl_layers} that early SSL layers encode low-level acoustic and phonetic information, while higher layers become increasingly specialized for linguistic content, which are less relevant for phonation discrimination. Among the three foundation models, HuBERT performed best across all layers compared to wav2vec2 models. Between wav2vec2 models, wav2vec2-LARGE performed better than wav2vec2-BASE.

Table \ref{table:model_accuracy} presents the results for the best-performing layer of foundation model features (layer 0 for wav2vec2-BASE and wav2vec2-LARGE, and layer 5 for HuBERT, irrespective of the classifier), along with the result of best baseline feature (spectrogram) for comparison. It can be observed that HuBERT-based features demonstrate superior performance compared to wav2vec2 models with an absolute accuracy improvement of over 5\% for SVM classifier and nearly 9\% for XGB classifier. Between the wav2vec2 models for the best layer results, wav2vec2-BASE slighly performed better than wav2vec2-LARGE. We hypothesize that the larger model’s deeper representations may over-specialize toward speech recognition targets, leading to reduced sensitivity to subtle voice-quality cues present in sustained singing vowels. However, note that wav2vec2-LARGE performed better across all other layers compared to wav2vec2-BASE. In comparison to the best baseline spectrogram feature (79.9\% for SVM and 79.6\% for XGB), HuBERT showed an absolute accuracy improvement of 12-15\%, and wav2vec2 models showed an absolute accuracy improvement of over 3-10\%.

Overall, SVM outperformed XGB for both conventional features and foundation model features. SVM for the spectrogram feature yielded the highest accuracy among the conventional features (79.9\%). When employing the same classifier with pre-trained foundation model-based features, the HuBERT-based features exhibited superior performance (95.7\%). These results highlight the effectiveness of HuBERT-based features, as integrated into our \textbf{voice2mode} system, showcasing a significant absolute improvement of 12–15\% compared to the best-performing baseline spectrogram feature.

Figure \ref{fig:cm} shows the confusion matrices for the best performing baseline (spectrogram, shown on the left) and best performing layer of foundation model (HuBERT, shown on the right) in order to see the class-wise phonation mode accuracies. It can be observed that HuBERT features exhibit better ability in distinguishing between phonation modes that are closely related, whereas the spectrogram features yield greater confusion in such scenarios, such as in distinguishing breathy from normal phonation, and pressed from flow phonation. 

\section{Conclusion}
We introduced \textit{voice2mode}, a novel framework for singing-voice phonation-mode classification that leverages self-supervised speech models as feature extractors. By repurposing HuBERT and wav2vec2.0 embeddings, models never explicitly trained on singing, we achieve dramatically better performance than standard handcrafted features. HuBERT features yielded the highest accuracy ($\approx$95\% with SVM), an absolute jump of about 12–15\% over the best baseline \cite{brandner2023classification}. Our layer-wise analysis indicates that the early transformer layers, rich in general acoustic information, are most useful for this task. These findings suggest that even without singing-specific pre-training, speech-based foundation models can capture the acoustic signatures of vocal fold control.

The success of \textbf{voice2mode} opens new possibilities. It demonstrates that modern deep audio representations generalize beyond speech and can be exploited in Music Information Retrieval. Practically, our system could power intelligent singing training tools: for example, giving real-time feedback on phonation mode to students, aiding vocal pedagogy. In future work, we plan to explore fine-tuning the foundation models on larger singing datasets, and to extend the approach to other voice-quality tasks (e.g. emotion or style in singing). We also release our code to encourage reproducibility and further advances in singing voice analysis.

While the present study focuses on sustained vowels from a single soprano singer and on a simplified set of phonation labels, future research should investigate broader datasets encompassing diverse voice types, genres, and expressive contexts. Incorporating temporal dynamics and contextual modelling could also enhance performance for continuous singing passages. Ultimately, this work contributes to bridging the gap between speech and singing research, showing that self-supervised speech representations hold untapped potential for high-level music understanding and expressive voice analysis.

\bibliographystyle{IEEEbib}
\bibliography{references}

\end{document}